\documentclass[prl,twocolumn,superscriptaddress]{revtex4}
\usepackage{graphicx}

\begin{document}

\newcommand{\be}{\begin{equation}}
\newcommand{\ee}{\end{equation}}
\newcommand{\bq}{\begin{eqnarray}}
\newcommand{\eq}{\end{eqnarray}}
\newcommand{\bsq}{\begin{subequations}}
\newcommand{\esq}{\end{subequations}}
\newcommand{\bc}{\begin{center}}
\newcommand{\ec}{\end{center}}

\newcommand {\R}{{\mathcal R}}
\newcommand{\al}{\alpha}

\title{Topological Defects in Contracting Universes}

\author{P.P. Avelino}
\email[Electronic address: ]{pedro@astro.up.pt}
\affiliation{Centro de Astrof\'{\i}sica da Universidade do Porto, R. das
Estrelas s/n, 4150-762 Porto, Portugal}
\affiliation{Departamento de F\'\i sica da
Faculdade de Ci\^encias da Universidade do
Porto, Rua do Campo Alegre 687, 4169-007, Porto, Portugal}
\author{C.J.A.P. Martins}
\email[Electronic address: ]{C.J.A.P.Martins@damtp.cam.ac.uk}
\affiliation{Centro de Astrof\'{\i}sica da Universidade do Porto, R. das
Estrelas s/n, 4150-762 Porto, Portugal}
\affiliation{Department of Applied Mathematics and Theoretical Physics,
Centre for Mathematical Sciences,\\ University of Cambridge,
Wilberforce Road, Cambridge CB3 0WA, United Kingdom}
\affiliation{Institut d'Astrophysique de Paris, 98 bis Boulevard Arago,
75014 Paris, France}
\author{C. Santos}
\affiliation{Centro de F\'\i sica da Universidade do Porto,
Rua do Campo Alegre 687, 4169-007, Porto, Portugal}
\email[Electronic address: ]{cssilva@fc.up.pt}
\affiliation{Departamento de F\'\i sica da
Faculdade de Ci\^encias da Universidade do
Porto, Rua do Campo Alegre 687, 4169-007, Porto, Portugal}
\author{E.P.S. Shellard}
\email[Electronic address: ]{E.P.S.Shellard@damtp.cam.ac.uk}
\affiliation{Department of Applied Mathematics and Theoretical Physics,
Centre for Mathematical Sciences,\\ University of Cambridge,
Wilberforce Road, Cambridge CB3 0WA, United Kingdom}

\date{24 August 2002; minor revisions 30 October 2002}

\begin{abstract}
We study the behaviour and consequences
of cosmic string networks in contracting universes.
They approximately
behave during the collapse phase as a radiation fluids.
Scaling solutions describing this are derived and tested
against high-resolution numerical simulations.
A string network in a contracting universe, together 
with the gravitational radiation it generates, 
can affect the dynamics of the universe both locally and 
globally, and be an important source of radiation, entropy 
and inhomogeneity. We discuss possible implications
for bouncing and cyclic models.
\end{abstract}
\preprint{DAMTP-2002-135}
\maketitle

\section{\label{intr}Introduction}

Cosmological scenarios involving oscillating or cyclic universes
have been know for a long time \cite{Tolman}. 
Recent interest has been associated with a cyclic extension
of the ekpyrotic scenario \cite{Steinhardt}.
A related result was
the realization \cite{Kanekar,Peter1,Peter2}
that the presence of a scalar field seems to
be necessary to make cosmological scenarios with a bounce
observationally realistic. And if scalar fields are present,
we should contemplate the possibility of topological
defect formation \cite{Book}.
Here we study cosmic string evolution in a
collapsing universe, and discuss some implications of the
presence of cosmic strings
for bouncing universes. In a bouncing universe scenario the 
properties of the universe in the expanding phase depend on 
physics happening in a previous collapsing phase.
Hence, if defects do exist in these models,
it is crucial to understand their evolution and consequences
in both the expanding and collapsing phases.
In particular, we expect that cosmic strings will become
ultra-relativistic, behaving approximately like a radiation
fluid. This means that a cosmic string network, both directly and through
the gravitational radiation emitted by its loops, will
soon become a significant source of entropy (and inhomogeneity),
making it a further problem for
cyclic universes if a suitable and efficient mechanism for diluting the
entropy is not available. A more detailed analysis can be
found in \cite{Cyclic}.

\section{\label{stringev}Cosmic string evolution}

The world history of a cosmic string can be represented by a
two-dimensional surface in space-time, obeying the usual Goto-Nambu action,
from which it is easy \cite{Book} to derive the microscopic string
equations of motion.
Consider for a start the evolution of a circular cosmic string loop in a 
cyclic universe. 
Simple analytic arguments show that a loop whose initial
radius is much smaller than the Hubble radius will oscillate 
freely with a constant invariant loop radius and an average velocity
${\bar v}=1/{\sqrt 2}$. (Note that we are assuming units in which $c=\hbar=1$.) 
On the other hand, once the collapse
phase begins, we will eventually get to a stage in which the physical
loop radius becomes comparable to the Hubble radius $ar\sim H^{-1}$ 
and then gets above it.
In this regime the loop velocity is typically driven towards unity 
$v\rightarrow 1$ and it straightforward to show that the 
invariant loop length (which is proportional 
to the energy of the loop) grows as $R \propto a^{-1}$ and the Lorentz factor as
$\gamma\propto a^{-2}$.  Despite its growing energy $R$, the actual physical 
loop radius $ar = R/\gamma \rightarrow a$,
so the loop shrinks with the scale factor
and inexorably follows the collapse into 
final big crunch singularity. 

Importantly, this relativistic final 
state for a loop in a collapsing universe is generic and quite 
different to the initial condition
usually assumed for super-horizon loops in the expanding phase. There, 
loops begin with a vanishing velocity which only becomes significant when
the loop falls below the Hubble radius.  Such evolution cannot be reproduced in 
reverse during the collapsing phase without fine-tuning the velocity as the loop 
crosses outside the Hubble radius.  This simple fact introduces a 
fundamental time asymmetry for string evolution in a cyclic universe (and 
for all other defects). 
The analytic expectations for our circular loop solution 
can easily be confirmed by a numerical study \cite{Cyclic}. 
              
Two complementary approaches are available
to study the evolution of a cosmic string network: one can resort to
large numerical simulations \cite{Bennett,Allen1,Moore},
or one can develop analytic
tools \cite{Kibble,Thesis,Wiggly} which provide an
averaged description of the basic properties
of the network. We shall use the best
motivated of these analytic models, the velocity-dependent one-scale (VOS)
model \cite{Martins1,Martins2,Thesis,Martins3}.
The VOS model describes the string dynamics in terms of two `thermodynamical' 
parameters: the string RMS velocity, $v_\infty$, 
and a single length scale---the string network is thus assumed to be
a Brownian random walk on large enough scales, with a correlation
length $L$. Hence one can simply relate it with the energy density in
long strings as $\rho_\infty=\mu/L^2$,
where $\mu$ is the string mass per unit length.  Note that the commonly used
`correlation length' $L$ is really a measure of the invariant string
length or energy, rather than the typical  
curvature radius of the strings.  By including the appropriate 
Lorentz factor $\gamma_\infty = (1-v_\infty^2)^{-1/2}$ for the long
strings, we can denote the physical correlation length  
by $L_{\rm phys} = L\gamma_\infty^{1/2}$.
With this assumption 
the VOS model has one phenomenological parameter ${\tilde c}$,
commonly called the loop chopping efficiency, which describes the
rate of energy transfer from the long-string network to loops.
The evolution equations then take the following form \cite{Martins1,Martins2}
\bq
&&2\frac{dL}{dt} = 2(1+ v_\infty^2)H L + {\tilde c} v_\infty
+8{\tilde \Gamma}G\mu v^6_\infty\label{50},\\
&&\frac{d v_\infty}{dt} = (1-v^2_\infty)\left(\frac{k(v_\infty)}{L}
-2 H v_\infty \right) \,.
\label{51}
\eq
The final term in the evolution equation for the correlation length
describes the effect of gravitational back-reaction.
We are not including in either equation additional
terms arising from friction due to particle scattering \cite{Martins3},
which could conceivably be important during the final stages of collapse.
We shall return to this point below.
Here $k(v_\infty)$ is the momentum parameter, which is thoroughly
discussed in \cite{Martins3}. Notice that this is
positive for $0<v^2<1/2$ and negative for $1/2<v^2<1$.

When the contraction phase starts and the Hubble parameter becomes
negative the velocity will tend to increase:
as in the simple case of the circular loop,
the string velocity will gradually tend towards unity. In this approximation,
and neglecting for the moment the loop production and gravitational
back-reaction terms, the evolution
equation for the correlation length easily yields $L\propto a^2$. Note that
this is the same overall scaling law for the string network as that in a 
radiation-dominated expanding universe; the string network 
effectively behaves like a radiation component.
In terms of the physical correlation length, $L_{\rm phys} \propto a$, as 
if strings were being conformally contracted (except for their rapidly 
growing velocities).
However, there are several factors that must be considered which complicate
this simple state of affairs. First, there is the issue of
loop production. Under the above assumptions on velocity, but putting the
loop production term back in the correlation length equation, one finds
the following approximate solution in the radiation and matter eras,
\be
L_{rad}=\left(L_{max}-\frac{{\tilde c}}{2}\ln{a}\right)a^2\,
\label{asymprad}
\ee
\be
L_{mat}=\left[L_{max}+\frac{{\tilde c}}{2}\left(a^{-1/2}-1\right)\right]a^2\,
\label{asympmat}
\ee
where $L_{max}$ is the string correlation length at the
time of maximal size of the universe, and the scale factor at that time
was chosen to be unity (so the logarithm term in the first case is positive).
Hence if ${\tilde c}$ remains constant (or is
slowly varying), asymptotically the scale factor dependent terms will
dominate, so that $L\propto a^2 \ln{a}$ in the radiation era,
and $L\propto a^{3/2}$ in the matter era. The latter
is also the scaling law for the correlation length in the
matter-dominated, expanding universe. This highlights the
different roles played by loop production in the scaling behaviour of
a cosmic string network in the radiation and matter eras \cite{Kibble,Thesis}.
A strong argument can be made, however, for a relativistic correction
to the loop production term. In the simplest form of the 
VOS model there is an identification between the correlation 
length, $L$, and the physical distance $L_{\rm phys}$ which a string segment 
is expected to travel before encountering another segment of the same size 
forming a loop in the process. However, taking into account the Lorentz factor
in the physical correlation length, one would expect \cite{Cyclic}
that ${\tilde c} \propto \gamma_\infty^{-1/2}$ thus 
driving ${\tilde c}$ 
rapidly towards zero and asymptotically yielding our simple solution 
$L \propto a^2$ both in the 
radiation and matter eras. 
Of course, during re-collapse we expect that 
${\tilde c}$ will depend on a number of other properties of the 
string network such as the enhanced build-up of small scale structure due to
the contraction.  Eventually, however, the Hubble radius will fall below 
even the length scale of wiggles on the string after which our asymptotic 
solution $L\propto a^2$ should be valid.  In what follows, we shall consider
the two well-motivated cases,  first,  ${\tilde c}=\hbox{const.}\ne 0$
initially and, secondly, ${\tilde c}=0$ the probable asymptotic case.
Further supporting evidence for this behaviour is discussed in 
\cite{Cyclic}.

Returning to our analytic solutions for the constant loop
production case (\ref{asymprad}) and (\ref{asympmat}), we can use 
the velocity equation to find an
approximate, implicit solution
\be
1-v^2\propto a^4\exp{\left[\frac{2k(v)}{\lambda}\int
\frac{a^{1/\lambda}da}{L(a)}\right]}\,,
\label{asympv}
\ee
where $\lambda=1$ in the radiation era and $\lambda=2$ in the matter era.
Substituting (\ref{asymprad}-\ref{asympmat}) one respectively obtains
\be
1-v^2_{rad}\propto a^4\left(-\ln a\right)^{4k(v)/{\tilde c}}\\,
\label{asympvvrad}
\ee
\be
1-v^2_{mat}\propto a^{4+2 k(v)/{\tilde c}}\,.
\label{asympvvmat}
\ee
Hence in the limit where
$v\to 1$ and therefore $k\to 0$ the asymptotic solution would have the
form $\gamma^{-2}\propto (1-v^2)\propto a^4$.
The momentum corrections, which phenomenologically
account for the existence of small scale structures on the strings,
imply that convergence will be slower than this.
These solutions will still hold when one includes the gravitational
back-reaction term \cite{Martins3}.
As a final caveat, it is also worth
emphasizing that the VOS model assumes that the long string network
has a Brownian distribution on large enough scales, which may not
be a realistic approximation in a closed, collapsing universe.
This point clearly deserves further investigation.

\begin{figure}
\includegraphics[width=3in,keepaspectratio]{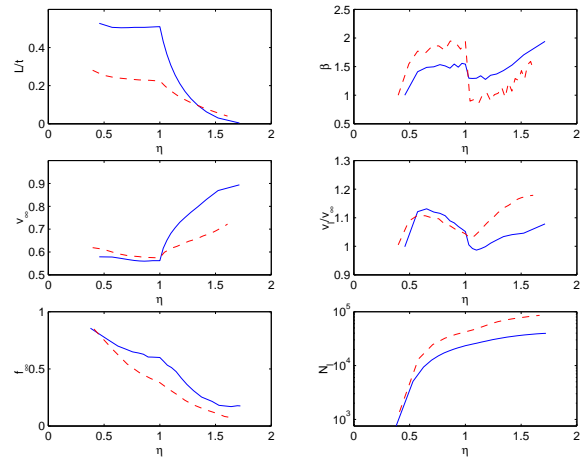}
\caption{\label{scalings}Cosmic string network properties
in the expansion and contraction phases, as a function of conformal time
(such that maximal expansion occurs at $\eta=1$ and
$\eta_{crunch}=2$). Solid is a simulation in the
matter era, dashed is radiation. Plotted are
the re-scaled correlation length $L/t$ (top left), the scaling law of
the correlation length
$L\propto a^\beta$ (top right), the velocity of the long string network
(middle left), the ratio of the loop and long string velocities (middle
right), the fraction of the string energy in the form of
long strings (bottom left) and the total number of loops (bottom
right).}
\end{figure}

As a test to the above solutions, we have performed a number of very
high resolution Goto-Nambu
simulations on the COSMOS supercomputer, using a
modified version of the Allen-Shellard string code \cite{Allen1}.
Further numerical details can be found in \cite{Cyclic}.
Our results are consistent with the existence of an
attractor solution of the type described above.
The result of two such simulations, for universes
filled with radiation and matter, is shown in Fig. \ref{scalings}.
During the expanding phase we confirm the usual linear scaling regimes
in the radiation and matter eras, respectively
\be
L_{exp,rad}\propto t\propto a^2,\qquad v_\infty=const.
\label{exprad}
\ee
\be
L_{exp,mat}\propto t\propto a^{3/2},\qquad v_\infty=const.
\label{expmat}
\ee
Once the contraction starts, these are modified:
the velocity starts increasing, and the scaling of the
correlation length with the scale factor also drops, being approximately
constant to begin with, and then rising slowly. One can
identify a transient scaling phase, valid in the period
$\eta\sim1.0-1.4$, where one approximately has
$L_{trans}\propto a$ in the radiation-dominated case, and
$L_{trans}\propto a^{5/4}$ in the matter era.
Unfortunately, the extremely
demanding requirements in terms of resolution of the simulation
do not currently allow us to run simulations with longer dynamic
range to establish beyond reasonable doubt
whether this scaling law approaches
$\beta=2$, as predicted above.  However, there are strong indications 
that the networks are evolving towards this asymptotic regime, as 
shown by the relatively rapid climb of the exponent in Fig.~\ref{scalings}.
It is clearly noticeable that the velocity rises much faster in
the matter era than in the radiation era. It is also interesting to
point out that during the collapse phase the loop and long string
velocities are noticeably different, and this difference (which is more
significant in the radiation than in the matter case) increases with time.
The plot also shows an apparent difference in the expanding phase, but this
is not significant: the initial lattice conditions tend to give different
velocities to small loops than to long strings and they start evolving,
and this difference is gradually erased.

We also notice that the network keeps chopping off loops
throughout the simulation, and that there is a dramatic increase
in the small scale structure of the network, particularly at later times.
Visually, the string network develops large  numbers of 
`knots', highly convoluted strings regions where the wiggly long strings 
have collapsed inhomogeneously.
These small scale features have proved to be difficult to evolve numerically,
and this in fact turns out to be the main limiting
factor at present preventing us
from running the simulations closer to the big crunch.
A comparison of our solutions (\ref{asymprad}-\ref{asympmat}) and
(\ref{asympvvrad}-\ref{asympvvmat}) with
the numerical simulations described above, produces 
a very good matching---see \cite{Cyclic}.
Finally it is also worth keeping in mind that any discussion of the
evolution of a cosmic string network with the present formalism is only
applicable while one is well below the Hagedorn temperature, at which 
the strings would `dissolve' in a reverse phase transition.
Discussions
of asymptotic regimes should be taken with some caution, since 
a cosmic string network will only survive the bounce intact if this happens
before the Hagedorn temperature is reached.

\section{\label{conseq}Discussion: cosmological consequences}

The overall density
of strings remains constant relative to the background
density $\bar\rho$ in both radiation and matter eras,
$\rho_\infty/{\bar\rho} = \sigma G\mu$,
with $\sigma _{\rm r}\approx 400$ and $\sigma_{\rm m}\approx 60$ respectively
\cite{Bennett,Allen1}.
During curvature domination or accelerated expansion, the string density grows
relative to the other matter as 
$\rho_\infty/\rho_{\rm m} \propto a$.   For GUT-scale strings 
with $G\mu \sim 10^{-6}$ this gives the interesting conclusion that today 
strings have a comparable energy density to the CMBR.  However, a realistic 
cyclic model will continue
to expand well beyond $t_0$, so the string density at maximum expansion 
will end up being much greater than the radiation density. 
In addition, the gravitational (or other) radiation produced through 
the continuous decay of the string network
evolves as $\rho_{\rm gr} \propto a^{-4}$.
It might appear that this contribution would become negligible during the 
matter era but in each Hubble time the strings lose about half their energy 
into gravitational radiation, so this background always remains comparable
to the string density $\rho_{\rm gr}\sim \rho_\infty$.
Now consider the collapsing phase in which the string network, like
the gravitational waves they have produced, begins to behave like radiation.  
Globally, the density of both the strings and the gravitational waves 
will grow as $a^{-4}$ and, together with any other radiation components,
they will eventually dominate over any nonrelativistic matter.  
In a realistic cyclic model, sufficiently massive strings and their
decay products 
will have a greater density than the microwave and neutrino backgrounds.
As the universe contracts, it will eventually reach a state in 
which the relativistic string network and/or their gravitational waves
dominate the global dynamics of the universe! This would lead to 
a dramatically different universe after it emerges from the next bounce.
Even lighter strings, which do not dominate the universe,  
would end up with a much greater density in the collapsing phase 
than they had previously during expansion.
If the universe went through a bounce, the energy density
in the cosmic strings and gravitational radiation produced by the network 
would be much greater after the bounce than before it.
Bounds on the string mass per unit length may be severely modified,
in addition to more general constraints on extra relativistic
fluids \cite{Bowen}.

Furthermore, unlike the uniform CMB, 
the energy density in both cosmic strings and gravitational
radiation will be very inhomogeneous.  In the collapsing regime,
an increasingly small fraction of Hubble regions will have a string
passing through them. Those that do will become string dominated since the 
string energy density in those regions will approximately evolve as
$\rho_\infty/{\bar \rho} \propto \gamma_\infty \propto a^{-2}$,
up to the corrections described above. For these regions
the assumption of a FRW background will cease to be valid at late times,
and the defects can make the universe
very inhomogeneous \cite{Inhomog} oranisotropic \cite{Fossils}.
Even Hubble regions without strings will have large fluctuations
in their gravitational radiation content.  For sufficiently 
massive strings, both of these effects
can survive the bounce to create large inhomogeneities 
in the next cycle.

A possible caveat to these solutions is dynamical friction (which
we have neglected so far). In the $\gamma >> 1$ limit,
one can estimate \cite{Cyclic} that strings loose all their
momentum due to this effects in one Hubble time when $G \mu\sim\gamma^{-2}$
(radiation era result) or $G \mu\sim \gamma^{-1}$ (matter era result).
However, this assumes a homogeneous and isotropic background, so need not
apply in our case. Moreover, the fact that a significant amount of momentum
will be transferred from the strings to the background will in itself
add to the anisotropies which naturally occur in our model. This will
be developed further elewehere \cite{Cyclic}.

We conclude that a cosmic string network will be a significant source of
radiation, entropy and inhomogeneity
which may be problematic in the cyclic context.
Some of the results described in this paper should also
be valid for other topological defects, in particular domain
walls. Conversely, if
direct evidence is found for the presence of topological defects
in the early universe, their existence
alone will impose constraints on the existence and characteristics of
previous phases of cosmological collapse.

\begin{acknowledgments}
C.M. and C.S. are funded by FCT (Portugal), under grants
FMRH/BPD/1600/2000 and BPD/22092/99 respectively.
Further support came from grant CERN/FIS/43737/2001.
This work was done in the context of the COSLAB network and
performed on COSMOS, the Origin3800 owned by the UK
Computational Cosmology Consortium, supported by Silicon Graphics/Cray
Research, HEFCE and PPARC.
\end{acknowledgments}

\bibliography{contracting}

\begin{thebibliography}{19}
\expandafter\ifx\csname natexlab\endcsname\relax\def\natexlab#1{#1}\fi
\expandafter\ifx\csname bibnamefont\endcsname\relax
  \def\bibnamefont#1{#1}\fi
\expandafter\ifx\csname bibfnamefont\endcsname\relax
  \def\bibfnamefont#1{#1}\fi
\expandafter\ifx\csname citenamefont\endcsname\relax
  \def\citenamefont#1{#1}\fi
\expandafter\ifx\csname url\endcsname\relax
  \def\url#1{\texttt{#1}}\fi
\expandafter\ifx\csname urlprefix\endcsname\relax\def\urlprefix{URL }\fi
\providecommand{\bibinfo}[2]{#2}
\providecommand{\eprint}[2][]{\url{#2}}

\bibitem[{\citenamefont{Tolman}(1934)}]{Tolman}
\bibinfo{author}{\bibfnamefont{R.~C.} \bibnamefont{Tolman}}
  (\bibinfo{year}{1934}), \bibinfo{note}{{ }Oxford, U.K.: Clarendon Press}.

\bibitem[{\citenamefont{Steinhardt and Turok}(2002)}]{Steinhardt}
\bibinfo{author}{\bibfnamefont{P.~J.} \bibnamefont{Steinhardt}}
  \bibnamefont{and} \bibinfo{author}{\bibfnamefont{N.}~\bibnamefont{Turok}},
  \bibinfo{journal}{Phys. Rev.} \textbf{\bibinfo{volume}{D65}},
  \bibinfo{pages}{126003} (\bibinfo{year}{2002}),
  \eprint[http://arXiv.org/abs]{hep-th/0111098}.

\bibitem[{\citenamefont{Kanekar et~al.}(2001)\citenamefont{Kanekar, Sahni, and
  Shtanov}}]{Kanekar}
\bibinfo{author}{\bibfnamefont{N.}~\bibnamefont{Kanekar}},
  \bibinfo{author}{\bibfnamefont{V.}~\bibnamefont{Sahni}}, \bibnamefont{and}
  \bibinfo{author}{\bibfnamefont{Y.}~\bibnamefont{Shtanov}},
  \bibinfo{journal}{Phys. Rev.} \textbf{\bibinfo{volume}{D63}},
  \bibinfo{pages}{083520} (\bibinfo{year}{2001}),
  \eprint[http://arXiv.org/abs]{astro-ph/0101448}.

\bibitem[{\citenamefont{Peter and Pinto-Neto}(2002{\natexlab{a}})}]{Peter1}
\bibinfo{author}{\bibfnamefont{P.}~\bibnamefont{Peter}} \bibnamefont{and}
  \bibinfo{author}{\bibfnamefont{N.}~\bibnamefont{Pinto-Neto}},
  \bibinfo{journal}{Phys. Rev.} \textbf{\bibinfo{volume}{D65}},
  \bibinfo{pages}{023513} (\bibinfo{year}{2002}{\natexlab{a}}),
  \eprint[http://arXiv.org/abs]{gr-qc/0109038}.

\bibitem[{\citenamefont{Peter and Pinto-Neto}(2002{\natexlab{b}})}]{Peter2}
\bibinfo{author}{\bibfnamefont{P.}~\bibnamefont{Peter}} \bibnamefont{and}
  \bibinfo{author}{\bibfnamefont{N.}~\bibnamefont{Pinto-Neto}},
  \bibinfo{journal}{Phys. Rev.} \textbf{\bibinfo{volume}{D66}},
  \bibinfo{pages}{063509} (\bibinfo{year}{2002}{\natexlab{b}}),
  \eprint[http://arXiv.org/abs]{hep-th/0203013}.

\bibitem[{\citenamefont{Vilenkin and Shellard}(1994)}]{Book}
\bibinfo{author}{\bibfnamefont{A.}~\bibnamefont{Vilenkin}} \bibnamefont{and}
  \bibinfo{author}{\bibfnamefont{E.~P.~S.} \bibnamefont{Shellard}}
  (\bibinfo{year}{1994}), \bibinfo{note}{{ }Cambridge, U.K.: Cambridge
  University Press}.

\bibitem[{\citenamefont{Avelino et~al.}(2002)\citenamefont{Avelino, Martins,
  Santos, and Shellard}}]{Cyclic}
\bibinfo{author}{\bibfnamefont{P.~P.} \bibnamefont{Avelino}},
  \bibinfo{author}{\bibfnamefont{C.~J. A.~P.} \bibnamefont{Martins}},
  \bibinfo{author}{\bibfnamefont{C.}~\bibnamefont{Santos}}, \bibnamefont{and}
  \bibinfo{author}{\bibfnamefont{E.~P.~S.} \bibnamefont{Shellard}}
  (\bibinfo{year}{2002}), \eprint[http://arXiv.org/abs]{astro-ph/0206287}.

\bibitem[{\citenamefont{Bennett and Bouchet}(1989)}]{Bennett}
\bibinfo{author}{\bibfnamefont{D.~P.} \bibnamefont{Bennett}} \bibnamefont{and}
  \bibinfo{author}{\bibfnamefont{F.~R.} \bibnamefont{Bouchet}},
  \bibinfo{journal}{Phys. Rev. Lett.} \textbf{\bibinfo{volume}{63}},
  \bibinfo{pages}{2776} (\bibinfo{year}{1989}).

\bibitem[{\citenamefont{Allen and Shellard}(1990)}]{Allen1}
\bibinfo{author}{\bibfnamefont{B.}~\bibnamefont{Allen}} \bibnamefont{and}
  \bibinfo{author}{\bibfnamefont{E.~P.~S.} \bibnamefont{Shellard}},
  \bibinfo{journal}{Phys. Rev. Lett.} \textbf{\bibinfo{volume}{64}},
  \bibinfo{pages}{119} (\bibinfo{year}{1990}).

\bibitem[{\citenamefont{Moore et~al.}(2002)\citenamefont{Moore, Shellard, and
  Martins}}]{Moore}
\bibinfo{author}{\bibfnamefont{J.~N.} \bibnamefont{Moore}},
  \bibinfo{author}{\bibfnamefont{E.~P.~S.} \bibnamefont{Shellard}},
  \bibnamefont{and} \bibinfo{author}{\bibfnamefont{C.~J. A.~P.}
  \bibnamefont{Martins}}, \bibinfo{journal}{Phys. Rev.}
  \textbf{\bibinfo{volume}{D65}}, \bibinfo{pages}{023503}
  (\bibinfo{year}{2002}), \eprint[http://arXiv.org/abs]{hep-ph/0107171}.

\bibitem[{\citenamefont{Kibble}(1985)}]{Kibble}
\bibinfo{author}{\bibfnamefont{T.~W.~B.} \bibnamefont{Kibble}},
  \bibinfo{journal}{Nucl. Phys.} \textbf{\bibinfo{volume}{B252}},
  \bibinfo{pages}{227} (\bibinfo{year}{1985}).

\bibitem[{\citenamefont{Martins}(1997)}]{Thesis}
\bibinfo{author}{\bibfnamefont{C.~J. A.~P.} \bibnamefont{Martins}}
  (\bibinfo{year}{1997}), \bibinfo{note}{{ }PhD Thesis, University of
  Cambridge}.

\bibitem[{\citenamefont{Martins}(2000)}]{Wiggly}
\bibinfo{author}{\bibfnamefont{C.~J. A.~P.} \bibnamefont{Martins}},
  \bibinfo{journal}{Nucl. Phys. Proc. Suppl.} \textbf{\bibinfo{volume}{81}},
  \bibinfo{pages}{361} (\bibinfo{year}{2000}).

\bibitem[{\citenamefont{Martins and Shellard}(1996{\natexlab{a}})}]{Martins1}
\bibinfo{author}{\bibfnamefont{C.~J. A.~P.} \bibnamefont{Martins}}
  \bibnamefont{and} \bibinfo{author}{\bibfnamefont{E.~P.~S.}
  \bibnamefont{Shellard}}, \bibinfo{journal}{Phys. Rev.}
  \textbf{\bibinfo{volume}{D53}}, \bibinfo{pages}{575}
  (\bibinfo{year}{1996}{\natexlab{a}}),
  \eprint[http://arXiv.org/abs]{hep-ph/9507335}.

\bibitem[{\citenamefont{Martins and Shellard}(1996{\natexlab{b}})}]{Martins2}
\bibinfo{author}{\bibfnamefont{C.~J. A.~P.} \bibnamefont{Martins}}
  \bibnamefont{and} \bibinfo{author}{\bibfnamefont{E.~P.~S.}
  \bibnamefont{Shellard}}, \bibinfo{journal}{Phys. Rev.}
  \textbf{\bibinfo{volume}{D54}}, \bibinfo{pages}{2535}
  (\bibinfo{year}{1996}{\natexlab{b}}),
  \eprint[http://arXiv.org/abs]{hep-ph/9602271}.

\bibitem[{\citenamefont{Martins and Shellard}(2002)}]{Martins3}
\bibinfo{author}{\bibfnamefont{C.~J. A.~P.} \bibnamefont{Martins}}
  \bibnamefont{and} \bibinfo{author}{\bibfnamefont{E.~P.~S.}
  \bibnamefont{Shellard}}, \bibinfo{journal}{Phys. Rev.}
  \textbf{\bibinfo{volume}{D65}}, \bibinfo{pages}{043514}
  (\bibinfo{year}{2002}), \eprint[http://arXiv.org/abs]{hep-ph/0003298}.

\bibitem[{\citenamefont{Bowen et~al.}(2002)\citenamefont{Bowen, Hansen,
  Melchiorri, Silk, and Trotta}}]{Bowen}
\bibinfo{author}{\bibfnamefont{R.}~\bibnamefont{Bowen}},
  \bibinfo{author}{\bibfnamefont{S.~H.} \bibnamefont{Hansen}},
  \bibinfo{author}{\bibfnamefont{A.}~\bibnamefont{Melchiorri}},
  \bibinfo{author}{\bibfnamefont{J.}~\bibnamefont{Silk}}, \bibnamefont{and}
  \bibinfo{author}{\bibfnamefont{R.}~\bibnamefont{Trotta}},
  \bibinfo{journal}{Mon. Not. Roy. Astron. Soc.}
  \textbf{\bibinfo{volume}{334}}, \bibinfo{pages}{760} (\bibinfo{year}{2002}),
  \eprint[http://arXiv.org/abs]{astro-ph/0110636}.

\bibitem[{\citenamefont{Avelino et~al.}(2001)\citenamefont{Avelino, Carvalho,
  Martins, and Oliveira}}]{Inhomog}
\bibinfo{author}{\bibfnamefont{P.~P.} \bibnamefont{Avelino}},
  \bibinfo{author}{\bibfnamefont{J.~P. M.~d.} \bibnamefont{Carvalho}},
  \bibinfo{author}{\bibfnamefont{C.~J. A.~P.} \bibnamefont{Martins}},
  \bibnamefont{and} \bibinfo{author}{\bibfnamefont{J.~C. R.~E.}
  \bibnamefont{Oliveira}}, \bibinfo{journal}{Phys. Lett.}
  \textbf{\bibinfo{volume}{B515}}, \bibinfo{pages}{148} (\bibinfo{year}{2001}),
  \eprint[http://arXiv.org/abs]{astro-ph/0004227}.

\bibitem[{\citenamefont{Avelino and Martins}(2000)}]{Fossils}
\bibinfo{author}{\bibfnamefont{P.~P.} \bibnamefont{Avelino}} \bibnamefont{and}
  \bibinfo{author}{\bibfnamefont{C.~J. A.~P.} \bibnamefont{Martins}},
  \bibinfo{journal}{Phys. Rev.} \textbf{\bibinfo{volume}{D62}},
  \bibinfo{pages}{103510} (\bibinfo{year}{2000}),
  \eprint[http://arXiv.org/abs]{astro-ph/0003231}.

\end{thebibliography}

\end{document}